\documentclass[12pt]{article}
\usepackage{a4wide,epsf}

\begin{document}

\begin{titlepage}
\renewcommand{\thefootnote}{\fnsymbol{footnote}}
\makebox[2cm]{}\\[-0.7in]
\begin{flushright}
\begin{tabular}{l}
CERN--TH/95--220\\
ULB--TH/95--11\\
hep-ph/9508359\\
August 1995
\end{tabular}
\end{flushright}
\vskip0.4cm
\begin{center}
{\Large\bf Phenomenological Evidence for the Gluon Content\\[4pt]
of $\mathbf{\eta}$ and $\mathbf{\eta'}$}

\vspace{1.5cm}

Patricia Ball$^1$,
J.-M.\ Fr\`{e}re$^2$\footnote{Directeur de recherches du FNRS}
and M. Tytgat$^2$\footnote{Aspirant du FNRS}

\vspace{1.5cm}

$^1${\em CERN, Theory Division, CH--1211 Gen\`{e}ve 23, Switzerland}\\[0.5cm]
$^2${\em Service de Physique Th\'{e}orique, Universit\'{e} Libre de
Bruxelles, CP 225,\\
B--1050 Bruxelles, Belgium}

\vspace{3cm}

{\bf Abstract:\\[5pt]}
\parbox[t]{\textwidth}{
We discuss the effect of $\eta$--$\eta'$ mixing and of axial current
anomalies on various decay processes. We describe the 2$\gamma$ decays of
$\eta$ and $\eta'$, the radiative decays of low-lying vector mesons as well
as the decays $\eta\to 3\pi$, $J/\psi\to\eta\gamma$ and $J/\psi\to\eta'\gamma$
in one unified framework, in which they are related to the electromagnetic and
strong axial current anomalies, respectively. We also discuss briefly the
enhancement of $\eta'$ in $D_s$ decays and suggest a possible relation to
gluon-mediated processes. The importance of the decays of the glueball
candidates $f_0(1500)$ and $ f_0(1590)$ into $\eta \eta'$ channels is also
stressed.}

\vfill
{\em Submitted to Phys.\ Lett.\ B.}
\end{center}
\end{titlepage}
\setcounter{footnote}{0}
\newpage

\noindent {\large\bf 1.} In this paper we study the effect of $\eta$--$\eta'$
mixing and of the axial current anomaly on various processes. We show that
a consistent picture arises for the radiative decays $\eta\to 2\gamma$ and
$\eta'\to 2\gamma$, for radiative decays of type $V\to\eta\gamma$,
$\eta'\to V\gamma$ of the lowest-lying vector mesons $V$, as well as for $\eta
\to 3\pi$ and $J/\psi\to\eta(\eta')\gamma$. Whereas for all the other decays
we follow closely the approach of Ref.\ \cite{AF}, our description of the
vector meson decays is based on their relation to the QED triangle anomaly.
In Sec.\ 2 we thus first test our approach for the simpler cases of the
decays $V\to\pi(K) \gamma$. In Sec.\ 3 we then establish notations for the
$\eta$--$\eta'$ system and determine the decay constants $f_0$ and $f_8$ as
functions of the mixing angle from the $2\gamma$ decays of $\eta$ and $\eta'$.
Section 4 is devoted to the comparison of our expectations for the widths of
$V\to\eta\gamma$, $\eta'\to V\gamma$ with experiment, whereas Sec.~5 deals
with the decay $\eta\to 3\pi$, which probes the light quark content of $\eta$.
In Sec.~6 we then investigate the gluon content of $\eta$ and $\eta'$, which
determines the decays $J/\psi\to\eta\gamma$ and $J/\psi\to\eta'\gamma$. It is
in this channel that our approach yields a
dependence of the involved matrix elements on the mixing angle that is
significantly different from other models. We briefly comment on this in
relation to the approach of Veneziano et al., Ref.\ \cite{venezia},
and close in Sec.\ 7 with some remarks on the important r\^{o}le of
anomalies in other gluon-enriched channels like $D_s\to\eta(\eta')X$ and
in glueball searches.

\bigskip

\noindent {\large\bf 2.} The radiative decays of the lowest-lying vector-meson
nonet are traditionally described in terms of magnetic moments of quarks, see
e.g.\ \cite{BT92}, or of unknown couplings related by SU$_\mathrm{
F}$(3) symmetry \cite{benayoun}. In Refs.\ \cite{chanowitz,SV81} the
width $\Gamma(\rho\to\pi\gamma)$ was related via SU$_\mathrm{F}$(3)
arguments and vector-meson dominance to the radiative width of $\pi$. In this
paper we relate the radiative decays of light vector mesons to light
pseudoscalars, $V\to P\gamma$ and $P\to V\gamma$, directly to the
anomaly of the $AVV$ triangle diagram, where $A$ stands for an axial-vector
and $V$ for a vector current. Our approach\footnote{A similar approach was
proposed in Ref.\ \cite{IS}.} both includes SU$_\mathrm{F}$(3) breaking
effects and fixes the vertex couplings $g_{VP\gamma}$ as defined
below\footnote{The behaviour of the anomalous contribution has been studied
quite extensively, in particular for large space-like momenta, and found to
fall off asymptotically like $1/-q^2$ \cite{spacelike}. The results
of this section show however that in the physical sector, and for
momenta up to the square of the $\phi$ mass, the full value of the anomaly
gives sensible results. It was important for us to check this fact in the
$\rho ,\omega, \pi, K $ sector, independently of further difficulties
associated with the strong anomaly.}.

Let us start by considering the correlation function
\begin{equation}
i\!\!\int\!\!d^4\!x\,e^{iq_2x}\,\langle\pi(q_1+q_2)|\,T\,
J^\mathrm{ EM}_\mu(x)J^{I=0(1)}_\nu(0)\,|0\rangle =
\epsilon_{\mu\nu\rho\sigma}q_1^\rho q_2^\sigma\,F_{\pi,I=0(1)}(q_1^2,q_2^2),
\end{equation}
with the currents
\begin{equation}
J^\mathrm{ EM}_\mu = \frac{2}{3}\,\bar u\gamma_\mu u - \frac{1}{3}\,
\bar d \gamma_\mu d - \frac{1}{3}\,\bar s \gamma_\mu s,\quad
J^{I=0,1}_\mu = \frac{1}{\sqrt{2}}\left( \bar u \gamma_\mu u \pm \bar d
\gamma_\mu d\right).
\end{equation}
The values of $F_{\pi,I}(0,0)$ are fixed by the QED triangle anomaly as
\begin{equation}
F_{\pi,I=1}(0,0) = \frac{1}{4\pi^2f_\pi},\qquad F_{\pi,I=0}(0,0) =
\frac{3}{4\pi^2 f_\pi},
\end{equation}
where the pion leptonic decay constant $f_\pi$ is defined as
\begin{equation}\label{eq:deffpi}
\langle\,0\,|\,\bar u \gamma_\mu\gamma_5 d\,|\,\pi^-\,\rangle = if_\pi p_\mu,
\end{equation}
so that $f_\pi = 0.132\,$GeV from $\pi\to\mu\nu_\mu$. For the analogously
defined kaon decay constant one finds
$f_K = 0.160\,$GeV. Using their analytic properties, we can express these
form factors by a dispersion relation in
the momentum of the isospin current\footnote{Note that the dispersion relation
needs no subtraction, and that the value $F_{\pi,I}(0,0)$ is unambiguously
fixed by the anomaly.}:
\begin{equation}\label{eq:spekrepr}
F_{\pi,I=1(0)}(0,0) = \frac{1}{\pi}\int\limits_{4(9)m_\pi^2}^\infty\!\!ds\,
\frac{1}{s}\,{\rm Im}\,F_{\pi,I=1(0)}(s,0).
\end{equation}
Saturating by the lowest-lying resonances we obtain
\begin{equation}\label{eq:xyz}
F_{\pi,I=1}(0,0) = \frac{f_\rho}{m_\rho}\,g_{\rho\pi\gamma}+\dots,\qquad
F_{\pi,I=0}(0,0) = \frac{f_\omega}{m_\omega}\,g_{\omega\pi\gamma}+\dots
\end{equation}
Here the dots stand for higher resonances and multi-particle contributions
to the correlation function. In the following we assume vector meson
dominance and thus neglect these contributions. Although this assumption may
be criticized (and actually in Ref.\ \cite{SV81} the continuum
contributions were modelled by the perturbative spectral function above some
threshold, but without taking into account the motion of the light quarks
in the pion), it yields a description of the data that is good enough for
our purposes\footnote{We are interested here in understanding how the gluon
anomaly affects decays involving $\eta$ and $\eta'$ and not so much in a
detailed fit of radiative vector meson decays.}.

The  $f_V$ in (\ref{eq:xyz}) are the vector mesons' leptonic decay constants
defined by
\begin{equation}
\langle\,0\,|\,J^V_\mu\,|\,V(p,\lambda)\,\rangle =
m_Vf_V\epsilon_\mu^{(\lambda)}(p).
\end{equation}
$\lambda$ denotes the helicity state of the meson. For ideal
mixing the relevant currents are $J_\mu^\rho = J_\mu^{I=1}$, $J_\mu^\omega =
J_\mu^{I=0}$ and $J_\mu^{\phi} = -\bar s \gamma_\mu s$. In the following we
include the deviation from ideal mixing by taking into account a mixing-angle
$\theta_V=40.3^\circ$ for $\phi$--$\omega$ mixing\footnote{This value follows
from the standard SU$_\mathrm{F}(3)$ breaking analysis; the ideal mixing angle
is given by $\tan\theta_V = 1/\sqrt{2}$, corresponding to
$\theta_V=35.3^\circ$.}.
The $f_V$ can be determined from the experimental decay rates \cite{partdata}
via
\begin{equation}
\Gamma(V\to e^+e^-) = c_V \pi\alpha^2\,\frac{f_V^2}{m_V}
\end{equation}
with $c_V = \{2/3,2/9\sin^2\theta_V,2/9\cos^2\theta_V\}$ for
$V=\{\rho^0,\omega,\phi\}$. The experimental results are
\begin{equation}
f_{\rho^0} = (216\pm5)\,{\rm MeV},\quad f_\omega = (174\pm 3)\,{\rm MeV},\quad
f_\phi = (254\pm 3)\,{\rm MeV}.
\end{equation}
The charged mesons decay constants can be obtained from $\tau\to V\nu_\tau$
via
\begin{equation}
\Gamma(\tau^-\to V\nu_\tau) = \frac{G_F^2|V_{ij}|^2}{16\pi}\,f_V^2m_\tau^3
\left(1-\frac{m_V^2}{m_\tau^2}\right)^2 \left(1+2\,\frac{m_V^2}{m_\tau^2}
\right),
\end{equation}
where $V_{ij}$ is the appropriate CKM matrix element. We find
\begin{equation}
f_{\rho^{\pm}} = (195 \pm 7)\,{\rm MeV},\qquad f_{K^{*\pm}} = (226 \pm 28)\,
{\rm MeV}.
\end{equation}
In view of the discrepancies between the measurements of the
$\rho$ decay constant from the charged and neutral sectors, we will
 use the average value $f_\rho = (205\pm 17)\,$MeV
in the following.
\begin{table}[tb]
\addtolength{\arraycolsep}{2pt}
\renewcommand{\arraystretch}{2.4}
$$
\begin{array}{ll|l@{\hspace*{2pt}}l|l}
P & V & \displaystyle g_{VP\gamma}\textrm{\ (th.)} & &
\displaystyle g_{VP\gamma}\textrm{\ (exp.)}\\
\hline\hline
\pi^\pm & \rho^\pm & \displaystyle\frac{m_\rho}{4f_\pi f_\rho\pi^2} & =(0.73
\pm 0.07)\,\textrm{GeV}^{-1} & (0.74\pm 0.04)\,\textrm{GeV}^{-1}\\
\pi^0 & \rho^0 & \displaystyle\frac{m_\rho}{4f_\pi f_\rho\pi^2} & =(0.73\pm
0.07)\,\textrm{GeV}^{-1} & (0.98 \pm 0.13)\,\textrm{GeV}^{-1}\\
\pi & \omega & \displaystyle\frac{3m_\omega}{4f_\pi f_\omega\pi^2}\,
\frac{\sin\theta_V+\sqrt{2}\,\cos\theta_V}{\sqrt{3}}&=(2.58\pm 0.04)\,
\textrm{GeV}^{-1} & (2.33\pm 0.07)\,\textrm{GeV}^{-1}\\
\pi & \phi & \displaystyle\frac{\sqrt{3}m_\phi}{4f_\pi f_\phi\pi^2}\left(
\sqrt{2}\sin\theta_V-\cos\theta_V\right) & =(0.20\pm0.01)\,\textrm{GeV}^{-1}&
(0.14\pm 0.01)\,\textrm{GeV}^{-1}\\
K^0 & \bar{K}^{0*} & \displaystyle\frac{m_{K^*}}{2f_{K^*}f_K\pi^2} &
=(1.25\pm 0.15)\,\textrm{GeV}^{-1} & (1.25\pm 0.05)\,\textrm{GeV}^{-1}\\
K^\pm & \bar{K}^{\pm *} & \displaystyle \frac{m_{K^*}}{4f_{K^*}f_K\pi^2} &
=(0.62\pm 0.08)\,\textrm{GeV}^{-1} & (0.84\pm 0.04)\,\textrm{GeV}^{-1}
\end{array}
$$
\vspace*{-0.5cm}
\addtolength{\arraycolsep}{-2pt}
\renewcommand{\arraystretch}{1}
\caption[]{Theoretical and experimental values of the on-shell $V$--$P$
electromagnetic vertex couplings defined in Eq.\ (\protect{\ref{eq:defg}}). For
$g_{VP\gamma}\,$(th.) we give only the experimental errors coming from the
decay constants $f_{P,V}$.}\label{tab:1}
\end{table}

Finally, we introduce vertex couplings $g_{VP\gamma}$, which are just the
on-shell V--P electromagnetic form factors:
\begin{equation}
\left.\langle \,P(p_P)\,|\,J^\mathrm{ EM}_\mu\,|\,
V(p_V,\lambda)\,\rangle\right|_{(p_V-p_P)^2=0} = -g_{VP\gamma}\,
\epsilon_{\mu\nu\rho\sigma}p_P^\nu p_V^\rho\epsilon^{(*)(\lambda)\sigma}_V\,.
\label{eq:defg}
\end{equation}
The amplitude of the decay $P\to V\gamma$ or $V\to P\gamma$, depending on
kinematics, is then obtained by contracting with the polarization vector of
the photon and multiplying by $\sqrt{4\pi\alpha}$. The decay rates read
\begin{equation}
\Gamma(P\to V\gamma) = \frac{\alpha}{8}\,g^2_{VP\gamma}\left(
\frac{m_P^2-m_V^2}{m_P}\right)^3,\quad
\Gamma(V\to P\gamma) = \frac{\alpha}{24}\,g^2_{VP\gamma}\left(
\frac{m_V^2-m_P^2}{m_V}\right)^3.
\end{equation}
In Table \ref{tab:1} we give both the theoretical and experimental values
of the couplings $g_{VP\gamma}$ for decays involving pions or kaons.
The agreement between theoretical and experimental values shows that the
ground states indeed dominate the spectral functions in (\ref{eq:spekrepr}),
with corrections of order 10\%. There are, however, three channels where
the deviation from the experimental value is larger. The couplings in the
$\rho$--$\pi$ channels should be equal, those in the $K^*$--$K$ channel differ
by a factor of two, which is not quite supported by the data. Since the
couplings
are related by Clebsch-Gordan coefficients, we see no possibility to
reconcile the data with our predictions and leave this ``anomaly'' as open
question to the experimentalists. For $\phi\to \pi\gamma$ the deviation is
more than 10\%. This decay, however, is in our approach completely due to
$\phi$--$\omega$ non-ideal mixing and thus strongly suppressed
compared to the other channels. It may also be influenced by other, usually
negligible mechanisms, in particular $\rho$--$\omega$ mixing, which we have
neglected here, so that we consider the agreement with the experimental
coupling as still satisfactory.

\bigskip

\noindent {\large\bf 3.}
Having gained some control over the radiative decays of vectors into $\pi$
and $K$, we now turn to the $\eta$--$\eta'$ system, which is our central point
of interest in this study. Numerous possibilities, of which we can only quote
a few \cite{mixed}, have been suggested to describe this system, with
or without explicit mixing with extra ``glueball'' states. The discussion of
all these approaches goes far beyond the scope of this letter, and we will
content ourselves with shortly commenting in Sec.~6 on the elegant approach by
Veneziano et al., Ref.\ \cite{venezia}. For the time being, however, we
focus on the more simple and pragmatic approach proposed in Ref.\ \cite{AF}.
Following the Particle Data Group conventions, which differ from the ones
originally used in Ref.\ \cite{AF}, we reproduce here the key formulas in the
current notation.
In terms of the mixing angle $\theta$ and their singlet and octet components
$\eta_0$ and $\eta_8$, respectively, the physical states are decomposed as
\begin{equation}
|\eta\rangle = |\eta_8\rangle\cos\theta- |\eta_0\rangle\sin\theta,\quad
|\eta'\rangle = |\eta_8\rangle\sin\theta+ |\eta_0\rangle\cos\theta.
\end{equation}
Traditional SU$_\mathrm{ F}(3)$ based analyses suggest either
$\theta\simeq -10^\circ$ or $\theta\simeq -23^\circ$ from the quadratic and
linear version, respectively, of the Gell--Mann-Okubo mass formula, whereas
we leave $\theta$ as an open parameter, at least for the time being.
Analogously to the pion decay constant $f_\pi$, Eq.\ (\ref{eq:deffpi}),
we define the decay constants\footnote{Note that by virtue of the ``hard''
non-conservation of $A^\mu_0$, $f_0$ is scale-dependent
as described in detail in the first reference of \cite{venezia};
 we will assume here that
the momentum dependence can be neglected.}
$f_8$ and $f_0$ as coupling of $\eta_8$ and $\eta_0$,
respectively, to the divergence of the relevant axial-vector currents
(where $f_8=f_\pi\neq f_0$ in the SU$_\mathrm{F}(3)$ limit):
\begin{eqnarray}
\partial_\mu A_8^\mu & = & \frac{2}{\sqrt{6}}\,\left( m_u \bar u i \gamma_5 u
+ m_d \bar d i \gamma_5 d - 2 m_s \bar s i \gamma_5 s\right),\nonumber\\
\partial_\mu A_0^\mu & = & \frac{2}{\sqrt{3}}\,\left( m_u \bar u i \gamma_5 u
+ m_d \bar d i \gamma_5 d + m_s \bar s i \gamma_5 s\right)+ \frac{1}{\sqrt{3}}
\,\frac{3}{4}\frac{\alpha_s}{\pi}\,G^A_{\mu\nu}\widetilde{G}^{A\mu\nu},
\label{eq:divergences}
\end{eqnarray}
where $G^A_{\mu\nu}$ is the gluonic field-strength tensor and
$\widetilde{G}^A_{\mu\nu}=\frac{1}{2}\epsilon_{\mu\nu\rho\sigma}
G^{A\rho\sigma}$ its dual.
Defining as in Ref.~\cite{AF} the interpolating fields
of $\eta$ and $\eta'$ as linear combinations of the
axial-vector divergences, we thus have\\[2pt]
\begin{equation}
\begin{array}[b]{r@{\ =\ }l@{\quad}r@{\ =\ }l}
\langle\,0\,|\,\partial_\mu A_8^\mu\,|\,\eta\,\rangle & m_\eta^2 f_8 \cos
\theta, & \langle\,0\,|\,\partial_\mu A_0^\mu\,|\,\eta\,\rangle & -m_\eta^2
f_0 \sin\theta,\\[13pt]
\langle\,0\,|\,\partial_\mu A_8^\mu\,|\,\eta'\,\rangle & m_{\eta'}^2 f_8 \sin
\theta, & \langle\,0\,|\,\partial_\mu A_0^\mu\,|\,\eta'\,\rangle &
m_{\eta'}^2 f_0 \cos\theta.\\[10pt]
\end{array}
\end{equation}

\begin{figure}
$$
\epsffile{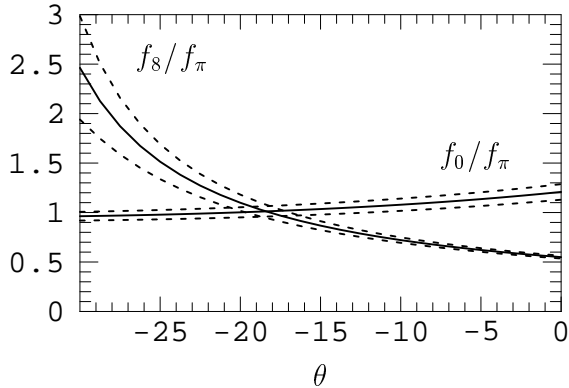}
$$
\vspace*{-1.0cm}
\caption[]{$f_0 /f_{\pi}$ and $f_8 /f_{\pi}$, determined from $\eta\to 2\gamma$
and $\eta'\to2\gamma$, as functions of the mixing angle.}\label{fig:f0f8}
\end{figure}

Our essential assumption is that the $u$ and $d$ quark masses can be
neglected in (\ref{eq:divergences}). This yields the following simple
expressions for the matrix element of the strong anomaly over the vacuum and
$\eta(\eta')$:
\begin{eqnarray}
\langle\,0\,|\frac{3}{4}\frac{\alpha_s}{\pi}\,G\widetilde{G}\,|\,\eta\,
\rangle & = & \sqrt{\frac{3}{2}}\,m_\eta^2\left(f_8\cos\theta-\sqrt{2}f_0
\sin\theta\right),\nonumber\\
\langle\,0\,|\frac{3}{4}\frac{\alpha_s}{\pi}\,G\widetilde{G}\,|\,\eta'\,
\rangle & = & \sqrt{\frac{3}{2}}\,m_{\eta'}^2\left(f_8\sin\theta+
\sqrt{2}f_0\cos\theta\right).
\label{eq:gluecontent}
\end{eqnarray}
We thus have three parameters to fix: $f_0$, $f_8$ and $\theta$. Two of them
can be determined from the radiative widths of $\eta$ and $\eta'$, which are
given by:
\begin{eqnarray}
\Gamma(\eta\to 2\gamma) & = & \frac{m_\eta^3}{96\pi^3}\,\alpha^2\left(
\frac{\cos\theta}{f_8}-\frac{2\sqrt{2}\sin\theta}{f_0}\right)^2,
\label{eq:form1}\\
\Gamma(\eta'\to 2\gamma) & = & \frac{m_{\eta'}^3}{96\pi^3}\,\alpha^2\left(
\frac{\sin\theta}{f_8}+\frac{2\sqrt{2}\cos\theta}{f_0}\right)^2.
\label{eq:form2}
\end{eqnarray}
As it was done in Ref.\ \cite{AF}, we solve these relations for the decay
constants and thus get $f_0$ and $f_8$ as functions of the mixing angle.
As experimental input we use \cite{partdata}
\begin{equation}\label{eq:exinput}
\Gamma(\eta\to 2\gamma) = (0.51\pm 0.026)\,{\rm keV},\quad
\Gamma(\eta'\to 2\gamma) = (4.53\pm 0.59)\,{\rm keV},
\end{equation}
where according to the suggestion in the full listings part of
the Review of Particle Properties (Ref.~\cite{partdata}, p.\ 1451)
we have only retained the more recent data.
In Fig.\ \ref{fig:f0f8}, $f_0 /f_{\pi}$ and $f_8 /f_{\pi}$ are plotted as
functions of the mixing angle (solid lines) together with their experimental
error (dashed lines).
We find thus $f_8/f_\pi \geq 1$ for $\theta$ less than $\approx -15^\circ$
as expected from chiral perturbation theory.
The precise value \cite{DHL85} $f_8/f_\pi=1.25$, however, corresponds
to a rather large\footnote{In fact, Ref.~\cite{mixangle} supplements the more
traditional approaches to the $2\gamma$ decays of $\eta$ and $\eta'$
by taking into account continuum contributions, which just cancel the
effect of the large SU$_\mathrm{F}$(3) breaking ratio $f_8/f_\pi=1.25$.
The result is then $\theta = -(17\pm 2)^\circ$ with the experimental input
data of Eq.~(\ref{eq:exinput}). As we shall
see in the following, this value is in excellent agreement with the results
we find from the investigation of other decay channels.}
mixing $\theta= -21.3^\circ$.

Having thus fixed $f_0$ and $f_8$, we continue in the next section with the
electromagnetic properties of $\eta$ and $\eta'$, i.e.\ the vector meson
decays, consider then in Sec.\ 5 their light quark matrix elements, to
finish in Sec.\ 6 with their glue content.
\begin{table}
\addtolength{\arraycolsep}{4pt}
\renewcommand{\arraystretch}{2.4}
$$
\begin{array}{ll|l|l}
P & V & \displaystyle g_{VP\gamma}\mbox{\ in units\ }\frac{m_V}{f_V\pi^2} &
\displaystyle g_{VP\gamma}\mathrm{\ (exp.)}\\[2pt]
\hline\hline
\eta & \rho &
\displaystyle\frac{\sqrt{3}}{4f_8}\,
\cos\theta - \sqrt{\frac{3}{8}}\,\frac{1}{f_0}\,\sin\theta
& (1.85\pm 0.34)\,\mathrm{GeV}^{-1}\\
\eta' & \rho & \displaystyle
\sqrt{\frac{3}{8}}\,\frac{1}{f_0}\,\cos\theta + \frac{\sqrt{3}}{4f_8}\,
\sin\theta
& (1.31\pm 0.12)\,\mathrm{GeV}^{-1}\\
\eta & \omega &
\displaystyle\frac{\cos\theta}{4f_8}\,(\sqrt{2}\cos\theta_V-\sin\theta_V)
-\frac{\sin\theta}{2\sqrt{2}f_0}\,\sin\theta_V
& (0.60\pm 0.15)\,\mathrm{GeV}^{-1}\\
\eta' & \omega & \displaystyle\frac{\cos\theta}{2\sqrt{2}f_0}\,\sin\theta_V
 + \frac{\sin\theta}{4f_8}\,(\sqrt{2}\cos\theta_V-\sin\theta_V)
& (0.45\pm 0.06)\,\mathrm{GeV}^{-1}\\
\eta & \phi & \displaystyle
-\frac{\cos\theta}{4f_8}\,(\cos\theta_V+\sqrt{2}\sin\theta_V)-
\frac{\sin\theta}{2\sqrt{2}f_0}\,\cos\theta_V &
(0.70\pm 0.03)\,\mathrm{GeV}^{-1}\\
\eta' & \phi & \displaystyle
\frac{\cos\theta}{2\sqrt{2}f_0}\,\cos\theta_V - \frac{\sin\theta}{4f_8}\,
(\cos\theta_V+\sqrt{2}\sin\theta_V) &
< 1.85\,\mathrm{GeV}^{-1}
\end{array}
$$
\vspace*{-0.5cm}
\addtolength{\arraycolsep}{-4pt}
\renewcommand{\arraystretch}{1}
\caption[]{Theoretical formulas and experimental values of the on-shell
$V$--$\eta,\eta'$ electromagnetic vertex couplings defined in
Eq.\ (\protect{\ref{eq:defg}}). In Fig.\ \protect{\ref{fig:fPV}} the
theoretical values are plotted as functions of $\theta$. We use
$\theta_V=40.3^\circ$ for the $\phi$--$\omega$ mixing angle. Experimental
values are taken from \protect{\cite{partdata}}.}\label{tab:2}
\end{table}
\begin{figure}
$$
\epsffile{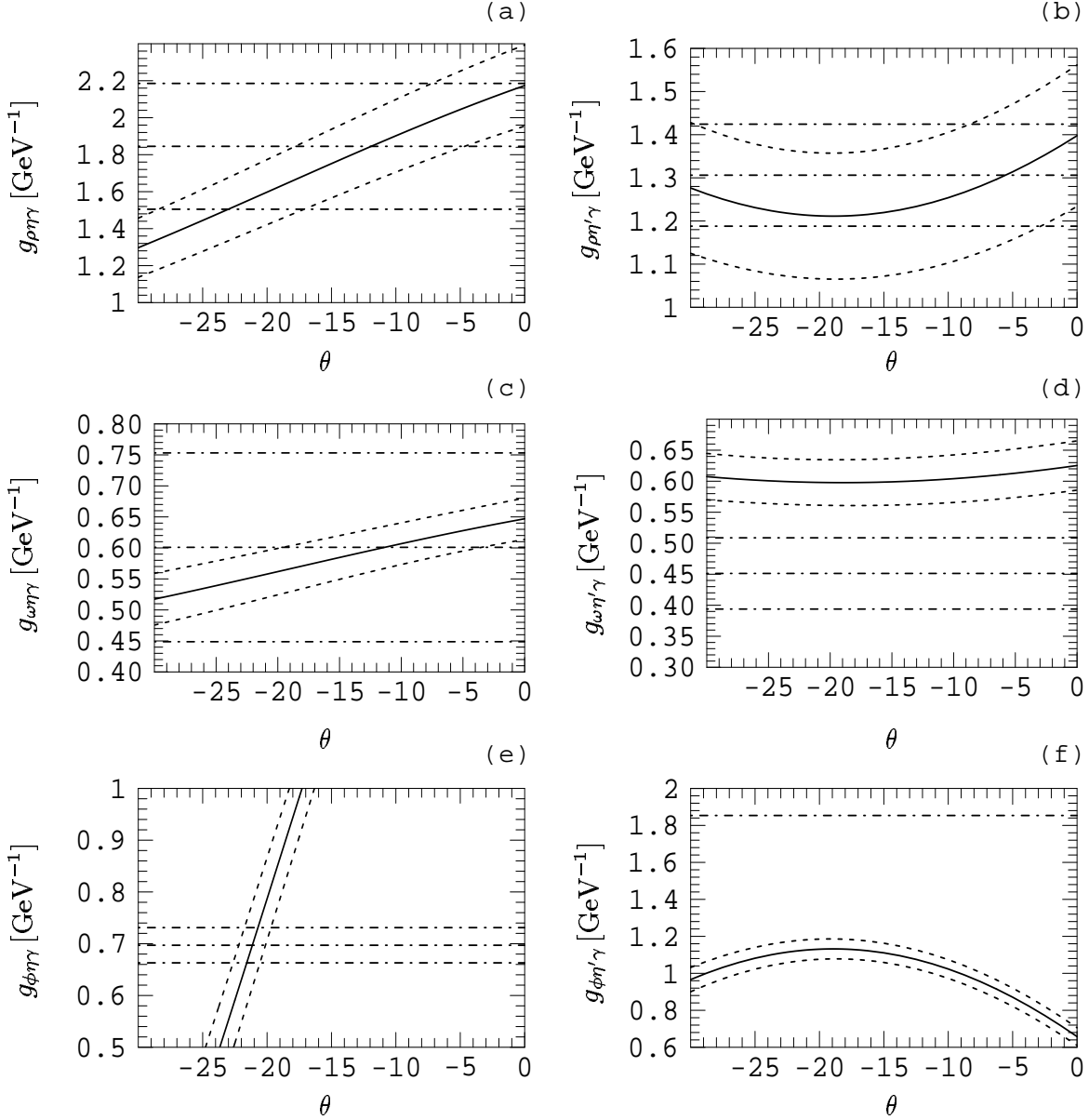}
$$
\caption[]{Theoretical and experimental values of the $V$--$\eta,\eta'$
electromagnetic couplings as functions of the pseudoscalar mixing angle.
Solid lines: theoretical
predictions according to Table \protect{\ref{tab:2}}. Short-dashed lines:
errors of the theoretical couplings coming from the experimental errors of
$f_0$ and $f_8$. Dashed-dotted lines: experimental couplings including
errors (for $g_{\phi\eta'\gamma}$, plotted in (f), there exists only an
experimental upper bound).}\label{fig:fPV}
\end{figure}

\bigskip

\noindent {\large\bf 4.} We are now in a position to deal with the radiative
vector meson decays involving $\eta$ or $\eta'$. In Table \ref{tab:2} we list
the decay channels together with the theoretical formulas for the couplings
$g_{VP\gamma}$ and their experimental values. In Fig.\ \ref{fig:fPV} we then
plot the theoretical and experimental values as functions of the pseudoscalar
mixing angle. Since in our approach continuum contributions are neglected,
the couplings should be slightly overestimated by about 10 to 15\%, as one
may also infer from the cleanest channel $\omega\to\pi\gamma$ investigated in
Sec.~2 (cf.\ Table~\ref{tab:1}). A look at Fig.~\ref{fig:fPV}(b) shows that
this is not the case for $\eta'\to\rho\gamma$. In that channel, however, we
expect a contamination of the experimental value from non-resonant $\pi\pi$
final states, which are produced by a completely different mechanism, which is
related to the box-anomaly diagram (cf.\ e.g.\ \cite{benabox}). For $\rho\to
\eta\gamma$ and $\omega\to\eta\gamma$, Figs.~\ref{fig:fPV}(a) and (c), the
experimental errors are too large to allow any serious restriction on $\theta$.
Within the error bars, however, our predictions agree with the experimental
values. $g_{\omega\eta'\gamma}$, on the other hand, is very sensitive to the
$\phi$--$\omega$ mixing angle; for instance, reducing $\theta_V$ by
1.5$^\circ$ brings the coupling within 20\% agreement with the experimental
data, i.e.\ just what we would expect.

As seen from the figures, the most stringent constraint on
$\theta$ currently stems from the decay $\phi \to \eta \gamma$, which, as
will turn out below, is only marginally consistent with the range of mixing
angles found from other processes (remember however that we use 1$\,\sigma$
experimental errors, and do not take into account theoretical uncertainties).
The issue should be clarified when experimental errors will eventually shrink,
and in particular we look forward to a measurement of the $\phi \to \eta'
\gamma$ mode, for which our prediction is only a factor 2 below the current
experimental upper bound.

\bigskip

\noindent {\large\bf 5.}
Let us now turn to the investigation of other matrix elements of $\eta$ and
$\eta'$ that can be probed in our approach.
We start with the light quark matrix elements,
which determine the decay $\eta \to 3 \pi$. As in Ref.~\cite{AF}
we find\footnote{Eq.\ (\ref{eq:ubaru}) differs slightly from the
corresponding Eq.\ (19) in Ref.\ \cite{AF}, where we had made the
simplification $f_0/f_8 \approx 1$.}:
\begin{equation}
\langle \, 0 \, | \, \bar u \gamma_5 u\, |\, \eta\, \rangle =
{}-\frac{1}{\sqrt{2}}\,\frac{f_0\cos\theta-\sqrt{2}f_8\sin\theta}{\sqrt{2}
f_0\cos\theta  + f_8 \sin\theta}\,\langle \, 0 \, | \, \bar s \gamma_5 s\, |\,
\eta\, \rangle \,,
\label{eq:ubaru}
\end{equation}
which yields the decay rate
\begin{equation}
\Gamma (\eta \to 3\pi^0) = \frac {\sqrt{3}}{4608\pi^2}\,
m_\eta^3 (m_{\eta} - 3 m_{\pi})^2 \delta_\eta
\left(\frac {m_d-m_u} {m_s} \right)^2 \frac{f_8^2\cos^2\!\theta}{f_\pi^6}
 \left( \frac {f_0\cos\theta   - \sqrt{2} f_8\sin\theta }
{\sqrt{2}f_0\cos\theta  + f_8 \sin\theta} \right)^2
\label{eq:threepi}
\end{equation}
with a kinematical factor $\delta_{\eta} = 0.86$. The most crucial ingredient
in that formula is the ratio of quark masses $r=(m_d-m_u)/m_s$. There exist
several possibilities to fix $r$ from next-to-leading order calculations in
chiral perturbation theory, as dicussed in
\cite{quarkmasses}, e.g. One of them is based essentially on the analysis of
$\eta\to3\pi$ in Ref.~\cite{famous}. Since in that study $\eta$ is necessarily
treated as Goldstone boson, the value of $r$ obtained that way is
inconsistent with our analysis. For similar reasons, we also refrain from
taking into account $r$ determined from $\Gamma(\psi'\to J/\psi\eta)/
\Gamma(\psi'\to J/\psi\pi)$. We thus rely on the determination of $r$ from
pseudoscalar mass relations, which according to Ref.~\cite{quarkmasses}
yield $r=0.030\pm0.005$.

\begin{figure}
$$
\epsffile{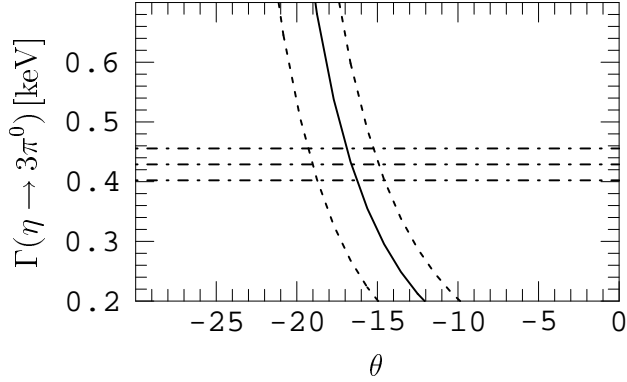}
$$
\vspace*{-1.0cm}
\caption[]{Theoretical (solid line) and experimental values (dashed-dotted
lines, including experimental error) of the decay rate $\Gamma(\eta
\to 3\pi^0)$ as function of the pseudoscalar mixing angle. The short dashes
give the theoretical error of the rate, which is dominated
by the error of the quark mass ratio.
}\label{fig:eta3pi}
\end{figure}

In Fig.~\ref{fig:eta3pi} we plot the decay rate as function of $\theta$.
Although the agreement with the experimental rate is very good for $\theta
\approx -17^\circ$, the theoretical errors are large and
dominated by the uncertainty in the quark mass ratio. Actually the true
theoretical error may be even larger and the rate be enhanced by final
state interactions \cite{quarkmasses,famous}.

In Ref.\ \cite{AF} we had also considered the decay $\eta' \to 3 \pi$.
We have however since argued \cite{castoldi} that this channel is considerably
more complex and also receives contributions from the leading
decay mode $\eta' \to \eta \pi \pi$ through $\eta$--$\pi$ mixing. While this
study is of great interest in itself and while a measurement
of the charged mode $\eta' \to \pi^+ \pi^- \pi^0 $ could shed
considerable light on the mechanism underlying $\eta' \to \eta \pi \pi$, it
is of little help in determining the value of $\theta$.

\bigskip

\noindent {\large\bf 6.}
We finally turn to the most important issue concerning the $\eta$--$\eta'$
system, namely the glue connection. Following the suggestion of Novikov et
al., Ref.\ \cite{NSVZ80}, the glue matrix elements are tested in
the ratio of the decay rates $\Gamma(J/\psi\to \eta(\eta')\gamma)$:
\begin{equation}
R_{J/\psi} \equiv \frac{\Gamma(J/\psi\to \eta'\gamma)}{\Gamma(J/\psi\to
\eta\gamma)} = \left|\frac{\langle\,0\,|\,G\widetilde{G}\,|\,\eta'\,
\rangle}{\langle\,0\,|\,G\widetilde{G}\,|\,\eta\,\rangle}\right|^2\,
\frac{\left(1-
m_{\eta'}^2/m_{J/\psi}^2\right)^3}{\left(1-m_{\eta}^2/m_{J/\psi}^2\right)^3}.
\label{eq:form3}
\end{equation}
Before turning to numerics, let us first comment shortly on the theoretical
accuracy of the above
equation. Actually the decay rates were calculated in exactly the same
approximation we used in deriving the $g_{VP\gamma}$ given in the tables,
namely dominance of the ground state and neglection of continuum contributions
to the dispersion relations. It was also pointed out in
Ref.\ \cite{BG} that the decay mechanism via the strong anomaly dominates
only due to the smallness of the $c$ quark mass and may be non-effective
already
in $\Upsilon$ decays, for which unfortunately no experimental data exist to
date. It is for these reasons that we prefer to consider the {\em ratio}
$R_{J/\psi}$ instead of the two decay rates separately: both radiative and
continuum corrections are expected to cancel to some extent in $R_{J/\psi}$.

\begin{figure}
$$
\epsffile{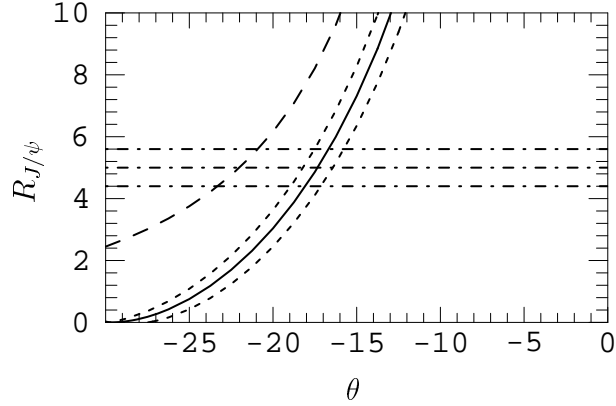}
$$
\vspace*{-1cm}
\caption[]{$R_{J/\psi}=\Gamma(J/\psi\to\eta'\gamma)/\Gamma(J/\psi\to\eta
\gamma)$ as function of the mixing angle $\theta$. Solid line: $R_{J/\psi}$
according to Eqs.\ (\protect{\ref{eq:gluecontent}}) and
(\protect{\ref{eq:form3}}) including experimental errors of
$f_0$ and $f_8$ (short dashes). Long-dashed line: the na\"\i\/ve prediction
(\protect{\ref{eq:naivemix}}). Dashed-dotted lines: experimental value with
errors.}\label{fig:Rjpsi}
\end{figure}
In Fig.\ \ref{fig:Rjpsi} we plot $R_{J/\psi}$ according to Eqs.\
(\ref{eq:gluecontent}) and (\ref{eq:form3}) and also the na\"\i\/ve prediction
\begin{equation}\label{eq:naivemix}
R_{J/\psi} = \mathrm{cot}^2\theta\,\,\frac{\left(1-
m_{\eta'}^2/m_{J/\psi}^2\right)^3}{\left(1-m_{\eta}^2/m_{J/\psi}^2\right)^3}
\end{equation}
as functions of the mixing angle.
As experimental input we use $R^{exp}_{J/\psi} = 5.0\pm 0.6$ \cite{partdata}.
Our prediction (\ref{eq:form3}) is rather sensitive to the mixing angle.
If we estimate the remaining theoretical uncertainty in (\ref{eq:form3}) as
$\sim 10\%$, we obtain $\theta=
-(17.3\pm 1.3)^\circ$. It is clearly seen, on the other hand, that the
 na\"\i\/ve curve is less consistent with the data and requires a very large
mixing angle. To understand the departure of our formula from
the na\"\i\/ve prediction, we briefly evoke the elegant, but heavier
formulation by Veneziano and collaborators \cite{venezia}.
These authors deal with a simplified model, where $\eta_8$--$\eta_0$ mixing
is neglected. To apply their approach to the physical situation requires
the introduction of {\em three}  initial fields (instead of two in our
approach):
\begin{equation}
  \phi^5_0 = \frac{1}{\sqrt{3}}\left(\bar u \gamma_5 u + \bar d
\gamma_5 d +  \bar s \gamma_5 s\right)\!,\quad
  \phi^5_8 = \frac{1}{\sqrt{6}}\left(\bar u \gamma_5 u + \bar d
\gamma_5 d -2 \bar s \gamma_5 s\right)\!,\quad
  Q =  \frac {\alpha_s} {8 \pi}\, G_{A \mu \nu} \tilde G^{A \mu \nu}.
\end{equation}
The definition of interpolating fields corresponding to physical particles
requires the orthonormalisation of these initial fields. While various
procedures are possible (like introducing some "glue" $Q$
in the definition
of $\eta'$, as we did above when using the total divergences as
interpolating fields), the choice of basis is essentially free,
and the authors of Ref.\ \cite{venezia} preferred to redefine the $Q$ field,
introducing $G$ as below\footnote{Note that the question if $G$ is to be
identified with any physical state, a glueball for instance, is left
open in Ref.~\cite{venezia}.}. This leads to (we follow closely the
notation of \cite{venezia}):
\begin{equation}
\phi_\eta = b_8 \cos\theta\,\phi^5_8 - b_0 \sin\theta\,\phi^5_0,\quad
\phi_{\eta'} = b_8 \sin\theta\,\phi^5_8 + b_0\cos\theta\,\phi^5_0,\quad
 G = Q - \frac {W_{\Theta \eta'}}{W_{\eta' \eta'}}\, \phi_{\eta'}
 - \frac{W_{\Theta \eta}}{W_{\eta \eta} }\, \phi_{\eta},\label{G}
\end{equation}
where the coefficients $b_i$ and $W_{ij}$ can be determined in terms of
two-point functions. Expressing the relevant Ward--Takahashi identities in
terms of the effective action and taking two derivatives with respect to
the gluon fields ($g$, with indices omitted below),
we obtain for the (non-anomalous) octet axial current:
\begin{equation}
 \Gamma_{\phi^5_8 g g} = b_8 \sin\theta\Gamma_{\eta' g g} +
 b_8\cos\theta  \Gamma_{\eta g g} + X  \Gamma_{G g g} =0
\end{equation}
with $X$ a computable coefficient.
Clearly, if the proper vertex between $G$ and two gluons could be neglected
(in the way the similar coupling between $G$ and two photons is legitimately
neglected in Ref.\ \cite{venezia}), we would recover the na\"\i\/ve result
(\ref{eq:naivemix}). This step is however not permitted since the proper vertex
between $G$ (whatever its physical realization) and two gluons is precisely
expected to be large. It should thus not be surprising that we have to depart
from Eq.\ (\ref{eq:naivemix}), which is difficult to reconcile with the data,
as already noted by Gilman and Kauffman in Ref.\ \cite{mixed}.
We will not dive deeper into this approach, except to mention that it is a
clear
generalization of ours (in our case the weight of glue in $\eta'$ is directly
fixed by our definition of its interpolating field, instead as resulting
from a further diagonalization between the three states).

\bigskip

\noindent {\large\bf 7.}
Having a rather comprehensive view of the $\eta$ and $\eta'$ system and
its connection with glue, we now address other promising channels for
future investigations, to wit $D_s$ decays. Here two processes compete to
produce $\eta(\eta') X$ final states: The first one is the
decay $c \to s$, leading to a $\bar s s $ pair, which hadronizes to $\eta$ or
$\eta'$, while the second one can be described as $c \bar s$ annihilation into
$W$ and two gluons. It is suppressed by the Zweig-rule, but may in the present
case gain importance through the large $\langle\,0\,|\,G\widetilde{G}\,|\,
\eta(\eta')\,\rangle$ matrix elements.
We should thus expect both an enhancement of the $(\eta +\eta')X$ modes
(unless the interference is destructive) and a large $\eta'/\eta$ ratio,
once the rates are corrected for phase-space and final state interactions.

If we ignore for the moment final state interactions, the
simplest way to evaluate a possible enhancement is by direct comparison to
similar $D_0$ decays. For instance the partial width for
$D_0 \to  K^- \rho^+$ (to which only the first diagram contributes) would, in
absence of the glue contribution, be expected to be of the same order (after
phase-space corrections) as the sum  $ D_s \to \eta \rho + D_s \to \eta' \rho$.
Instead we find:
\begin{equation}
\frac{\Gamma (D_s \to \eta' \rho) (p_{\eta'})^{-3} + \Gamma (D_s \to \eta
\rho)(p_{\eta})^{-3}} {\Gamma (D_0 \to K^- \rho) (p_{K})^{-3}} \left( \frac
{m_{D_s}} {m_{D_0}} \right) ^2 = 4.2 \pm 1.0,\label{eq:rhorate}
\end{equation}
while for the branching ratios we find
\begin{equation}
\frac {\Gamma (D_s \to \eta' \rho)} {\Gamma (D_s \to \eta \rho)}
\left(\frac {p_{\eta'}}{p_{\eta}}\right)^{-3} = 4.5 \pm 1.5,
\label{eq:rhoratio}
\end{equation}
which indicates both an enhancement of the decay rate and a large ratio for
these important decay modes (of the order of 10\% each).
The situation is unfortunately less clear in the $\pi$ associated modes,
where the comparison with $D_0$ now gives:
\begin{equation}
\frac{\Gamma (D_s \to \eta' \pi) (p_{\eta'})^{-1} + \Gamma (D_s \to \eta \pi)
(p_{\eta})^{-1}} {\Gamma (D_0 \to K^- \pi) (p_{K})^{-1}}  = 1.6 \pm 0.4,
\label{eq:pirate}
\end{equation}
while
\begin{equation}
\frac {\Gamma (D_s \to \eta' \pi)} {\Gamma (D_s \to \eta \pi)}
\left(\frac {p_{\eta'}}{p_{\eta}}\right)^{-1} = 3.0 \pm 1.1.\label{eq:piratio}
\end{equation}

More sophisticated analyses of non-leptonic $D_s$ decays were the
subject of several studies, e.g.\ \cite{DS,interesting,italians},
which, however, all relied on the na\"\i\/ve picture of $\eta$--$\eta'$ mixing
without taking into account the anomaly. In Ref.\ \cite{interesting} it was
pointed out that in order to account for the data the non-leptonic rates
involving $\eta'$ need to be enhanced by an extra factor of up to three when
compared with the corresponding $D_0$ decay amplitudes, even after including
final state interactions. And the global fit done in
Ref.~\cite{italians} for all non-leptonic $D$ and $D_s$ decays including final
state interactions, $W$ exchange and annihilation contributions fails to
explain the large branching ratios of $D_s$ into $\eta'\pi$ and $\eta'\rho$.
The enhancement needed is however well in line with our expectations from the
extra Zweig-suppressed (gluon-mediated) amplitudes.

Very recently, the CLEO collaboration has measured the
theoretically cleanest semileptonic channel and finds \cite{CLEO}
\begin{equation}
\frac{\Gamma(D_s\to\eta'e\nu)}{\Gamma(D_s\to\eta e\nu)} = 0.35\pm 0.09\pm 0.07.
\end{equation}
Defining the relevant formfactors as
\begin{equation}
\langle\,\eta\,|\,\bar s \gamma_\mu c\,|\,D_s\,\rangle = f_+^\eta(q^2)\,
(p_{D_s} + p_\eta)_\mu + f_-^\eta(q^2)\,(p_{D_s} - p_\eta)_\mu
\end{equation}
with $q = p_{D_s}-p_\eta$ and analogously $f_\pm^{\eta'}$, CLEO's
measurement yields
\begin{equation}
\frac{f_+^{\eta'}(0)}{f_+^\eta(0)} = 1.14\pm 0.17\pm 0.13,
\end{equation}
where we assumed monopole form
factors\footnote{We do not expect the anomalous
contribution to change drastically the $q^2$ dependence of the form factors,
which in $D\to K e \nu$ is a monopole experimentally.}
with a pole at $q^2 = m_{D_s^*}^2$.
The na\"\i\/ve mixing model, on the other hand, predicts
\begin{equation}
\frac{f_+^{\eta'}(q^2)}{f_+^\eta(q^2)} = \frac{\cos\theta-\sqrt{2}\sin
\theta}{\sin\theta+\sqrt{2}\cos\theta},
\end{equation}
which translates into $\theta = -(13.5\pm 4.7)^\circ$ and is only marginally
consistent with the determination from $\eta\to 2\gamma$ \cite{mixangle}.

As a temporary conclusion to this section, we would like to add that,
while this sector is clearly difficult to investigate both theoretically
and experimentally, it could prove very fruitful for our understanding of
the r\^ole of anomalies and the importance of gluons.
There is also the strong suggestion that $D_s$ decays, not unlike
the $J/\psi$ radiative decays, constitute a "glue-rich" channel,
where search for possible glueball states should be considered. We hope
to come back to that issue in future work.

Let us finally remark that apart from understanding the
structure of the $\eta$ and $\eta'$ particles, the size of their connection
to glue is of importance to investigate possible glueball candidates
\cite{gerstein}. In particular, the candidates $f_0(1500)$ from
crystal barrel \cite{crystal} and $f_0(1590)$ \cite{gams} are seen to decay
more frequently into $\eta \eta'$ than into $\eta \eta$ (after
phase-space correction), a ratio which we expect to be given by
\begin{equation}
\frac{\Gamma(f_0\to \eta'\eta)}{\Gamma(f_0 \to \eta\eta)} =
\left|\frac{\langle\,0\,|\,G\widetilde{G}\,|\,\eta'\,\rangle}{\langle\,0\,|\,
G\widetilde{G}\,|\,\eta\,\rangle}\right|^2\,\frac{p^{CM}_{\eta\eta'}}
{p^{CM}_{\eta\eta}}\,,\label{eq:gerstein}
\end{equation}
where $p^{CM}_{\eta \eta(')}$ are the respective centre of mass momenta.
This relation is however highly momentum dependent, since we are close to
the threshold, and obviously needs to be integrated over the particle width.

In conclusion, the above study gives a reasonably consistent description of
the $\eta$--$\eta'$ system for values of the mixing angle $\theta$ between
$-20$ and $-17$ degrees, taking into account anomalies, as tested from
various channels, respectively depicting eletromagnetic properties,
light quark and gluon content. Our approach will be tested by
a general improvement of the experimental
measurements, including the $V \to P \gamma$ and $P \to V \gamma$ rates,
and in particular by the observation of $\phi \to \eta' \gamma$.
The investigation of ``glue-rich'' channels, namely the traditional
radiative decays of $J/\psi$, but also, as advocated here,
the $D_s$ decays, will also yield a better understanding of this system
and at the same time help in the search for glueballs, which might
solve the puzzle.

\bigskip

\noindent{\bf Acknowledgements:} We thank C. Amsler, M. Benayoun,
X.Y.\ Pham, M. Shifman and G.~Veneziano for interesting discussions. P.B.\
gratefully acknowledges the hospitality of the LPTHE of Universit\'{e} Paris
VII, where part of this work was done. J.-M.F.\ also acknowledges discussions
with I.I.\ Bigi. This work was supported in part by the European Network
Flavourdynamics (ref.\ chrx-ct93-0132) and Nato grant (ref.\ CRG920611).

\end{document}